\documentclass{elsart}
\usepackage{graphicx}
\usepackage{hyperref}

\begin{document}

\begin{frontmatter}

\title{Search for Correlations between HiRes Stereo Events and 
Active Galactic Nuclei}

\author[utah]{R.~U. Abbasi},
\author[utah]{T.~Abu-Zayyad},
\author[utah]{M.~Allen},
\author[lanl]{J.~F.~Amman},
\author[utah]{G.~Archbold},
\author[utah]{K.~Belov},
\author[montana]{J.~W.~Belz},
\author[utah]{S.~Y.~BenZvi},
\author[rutgers]{D.~R.~Bergman},
\author[utah]{S.~A.~Blake},
\author[columbia]{J.~H.~Boyer},
\author[utah]{O.~A.~Brusova},
\author[utah]{G.~W.~Burt},
\author[utah]{C.~Cannon},
\author[utah]{Z.~Cao},
\author[utah]{W.~Deng},
\author[utah]{Y.~Fedorova},
\author[utah]{J.~Findlay},
\author[columbia]{C.~B.~Finley},
\author[utah]{R.~C.~Gray},
\author[utah]{W.~F.~Hanlon},
\author[lanl]{C.~M.~Hoffman},
\author[lanl]{M.~H.~Holzscheiter},
\author[rutgers]{G.~Hughes},
\author[lanl]{P.~H\"{u}ntemeyer},
\author[rutgers]{D.~Ivanov},
\author[utah]{B.~F~Jones},
\author[utah]{C.~C.~H.~Jui},
\author[utah]{K.~Kim},
\author[montana]{M.~A.~Kirn},
\author[columbia]{B.~C.~Knapp},
\author[utah]{E.~C.~Loh},
\author[utah]{M.~M.~Maestas},
\author[tokyo]{N.~Manago},
\author[columbia]{E.~J.~Mannel},
\author[lanl]{L.~J.~Marek},
\author[utah]{K.~Martens},
\author[utah]{J.~N.~Matthews},
\author[utah]{S.~A.~Moore},
\author[columbia]{A.~O'Neill},
\author[lanl]{C.~A.~Painter},
\author[rutgers]{L.~Perera},
\author[utah]{K.~Reil},
\author[utah]{R.~Riehle},
\author[unm]{M.~D.~Roberts},
\author[utah]{D.~Rodriguez}
\author[tokyo]{N.~Sasaki},
\author[rutgers]{S.~R.~Schnetzer},
\author[rutgers]{L.~M.~Scott\corauthref{cor}},
\corauth[cor]{Corresponding author.}
\ead{lscott@physics.rutgers.edu}
\author[columbia]{M.~Seman},
\author[lanl]{G.~Sinnis},
\author[utah]{J.~D.~Smith},
\author[utah]{R.~Snow},
\author[utah]{P.~Sokolsky},
\author[columbia]{C.~Song},
\author[utah]{R.~W.~Springer},
\author[utah]{B.~T.~Stokes},
\author[rutgers]{S.~R.~Stratton},
\author[utah]{J.~R.~Thomas},
\author[utah]{S.~B.~Thomas},
\author[rutgers]{G.~B.~Thomson},
\author[lanl]{D.~Tupa},
\author[utah]{L.~R.~Wiencke},
\author[rutgers]{A.~Zech},
\author[columbia]{X.~Zhang} \\
(The High Resolution Fly's Eye Collaboration)


\address[utah]{University of Utah, Department of Physics, Salt Lake City, UT 84112, USA}
\address[lanl]{Los Alamos National Laboratory, Los Alamos, NM 87545, USA}
\address[montana]{Montana State University, Department of Physics, Bozeman, MT 59812, USA}
\address[columbia]{Columbia University, Department of Physics and Nevis Laboratory, New York, NY 10027, USA}
\address[rutgers]{Rutgers --- the State University of New Jersey, Piscataway, NJ 08854, USA}
\address[tokyo]{University of Tokyo, Institute for Cosmic Ray Research, Kashiwa City, Chiba 277-8582, Japan}
\address[unm]{University of New Mexico, Department of Physics and Astronomy, Albuquerque, NM 87131, USA}

\title{}
\author{}

\begin{abstract}

We have searched for correlations between the pointing directions of ultrahigh
energy cosmic rays observed by the High Resolution Fly's Eye experiment and 
Active Galactic Nuclei (AGN) visible from its northern hemisphere location.
No correlations, other than random correlations, have been found.
We report our results using search parameters prescribed by the Pierre Auger
collaboration.
Using these parameters, the Auger collaboration concludes that a positive 
correlation exists for sources visible to their southern hemisphere location.  
We also describe results using two methods for determining the chance 
probability of correlations: one in which a hypothesis is formed from scanning 
one half of the data and tested on the second half, and another which involves 
a scan over the entire data set.
The most significant correlation found occurred with a chance probability of 
24\%.
\end{abstract}


\end{frontmatter}

\section{Introduction}

The search for the sources of the highest energy cosmic rays is an
important topic in physics today.
The energies of these cosmic rays exceed 100~EeV and the acceleration
mechanisms of the astrophysical objects responsible for these events 
remain unknown.
Anisotropy search methods such as those used in X- or $\gamma $-ray astronomy
are difficult to use due to deflections in the trajectories of these charged 
cosmic rays from Galactic and extragalactic magnetic fields.
For a galactic magnetic field strength of $\sim 3 \mu $G and coherence 
length of $\sim 1$~kpc, a 40 EeV cosmic ray should be deflected by two to 
three degrees over a distance of only a few kpc \cite{Dolag_JETP_2004}.

There are several reports on anisotropy by previous experiments.
An excess of events near the direction of the Galactic center has been
reported by the SUGAR and AGASA experiments 
\cite{Bellido_AstropartPhys_2001,Takeda_ApJ_1999}.
The Pierre Auger collaboration, however, has recently reported that they
have not seen any excess at that location \cite{Santos_ICRC2007}.
In addition, the Auger collaboration reported no significant excesses in 
any part of the southern hemisphere sky \cite{Mollerach_ICRC2007}.
Two reports of anisotropy have been found in the northern hemisphere sky.
A dip in the intensity of cosmic-ray events near the direction of the 
Galactic anticenter has been reported by both the AGASA and High Resolution
Fly's Eye (HiRes) experiments, but the significance is too low to claim an 
observation \cite{Ivanov_ICRC2007}.
Additionally the AGASA ``triplet'' is correlated with a HiRes high-energy 
event \cite{Abbasi_ApJ_2005b}.
These reports of anisotropy in the northern sky await confirmation or 
rejection by the Telescope Array experiment \cite{Fukushima_TA}.

Another method for searching for anisotropy is to search for correlations in 
pointing directions of cosmic rays with known astrophysical objects that might
be sources.
In these cases, a small event sample that shows no excess over the expected
background can, nevertheless, exhibit correlations with \emph{a priori}
candidate sources, adding up to a statistically significant signal.
Past searches have found correlations with BL Lacertae objects; BL Lacs
are a class of AGN with a jet pointing toward the Earth, and are plausible 
candidates for cosmic-ray sources.
Correlations have been found with data from the AGASA, HiRes and Yakutsk 
experiments, all in the northern hemisphere \cite{Tinyakov.Tkachev_JETP_2001}.
The Auger collaboration has searched for correlations with BL Lac objects
in the southern hemisphere but has found nothing significant
\cite{Harari_ICRC2007}.
Again the northern hemisphere correlations await confirmation by the Telescope
Array experiment.

There have been speculations that Active Galactic Nuclei (AGN) may contain 
acceleration regions of the appropriate size and magnetic field strength to 
accelerate nuclei to the highest energies 
\cite{Hillas_ARAA_1984,Berezinskii_1990}.
One should therefore expect the brightest and closest AGN to produce the 
highest-energy cosmic ray events at Earth.
These events would also have suffered the smallest deflections due to the 
intervening magnetic fields and would point back, most directly, to these AGN.
The large number of identified AGN make them interesting candidates for 
studying possible correlations with ultrahigh energy cosmic rays.
Three ideal parameters for determining correlations between cosmic rays and 
AGN are the maximum difference in angle between the cosmic-ray pointing 
direction and the AGN $\theta_{max}$, the minimum cosmic-ray energy $E_{min}$,
and the maximum AGN redshift $z_{max}$.

The Pierre Auger Collaboration have reported a search of two independent sets
of their data for correlations with cosmic rays with AGN.
They scanned their first data set and found that the most significant 
correlation occurs for cosmic rays with parameters 
($\theta_{max}$, $E_{min}$, $z_{max}$) = ($3.1^\circ$, 56~EeV, 0.018).
With these selection criteria, they find 12 pairings with AGN from 15 events
in the first data set.
In the second data set, they find 8 pairings from 13 events 
and a corresponding chance probability of 0.0017
\cite{Auger_AGN_2007,Auger_AGN_2007_full}.

The HiRes experiment collected data from 1997 to 2006, operating two 
fluorescence detectors located atop desert mountains separated by 
12.6~km in west-central Utah.
The HiRes data have been analyzed monocularly, using the data from one 
detector at a time \cite{Abbasi_PRL_2008}, and stereoscopically, using the data
from both detectors simultaneously \cite{Sokolsky_ICRC2007}.
The angular resolution is about $0.8^\circ$ in stereo mode.
The energy scales of the HiRes monocular and stereoscopic reconstructions 
agree.
Only stereo data were used in this analysis.
The stereo data, covering an energy range from $10^{17.4}$ to 
$10^{20.1}$~eV, consist of 6636 events.

The pointing directions of the stereo data extend from zenith to about 
$-32^\circ$ in declination (celestial coordinates).
The corresponding exposure of is dependent on right ascension due
to seasonal variations in the duty cycle of the detector.
The boundaries of regions of equal exposure are best described by

\begin{eqnarray}
  \label{eq:function}
  \delta = \left \{
  \begin{array}{ll}
     A + B \sin \left[ \frac{9}{10} \ \alpha \right] & \mbox{ (if $\alpha \leq 200^\circ$) }
     \\
     A + C \sin \left[ \frac{9}{8} \ (\alpha - 200^\circ) \right] 
         & \mbox{ (if $\alpha > 200^\circ$) }
  \end{array} \right.
\end{eqnarray}

\noindent
where $\delta$ and $\alpha$ are celestial declination and right ascension measured in 
degrees and $A$, $B$ and $C$ are fit parameters.
Table \ref{tab:exposure} gives values of $A$, $B$ and $C$ for plotting the boundaries
of the 10 bins of equal exposure shown in Figures~\ref{fig:auger_point} and 
\ref{fig:hires_point}.

\begin{table}[t]
  \caption{Parameters for the functions in Equation \ref{eq:function} that 
    give the coordinates (in celestial right ascension and declination) of the lower 
    boundaries of the 10 bins of equal exposure for the HiRes detector shown
    as the 10 lightest shaded regions in Figures~\ref{fig:auger_point} and 
    \ref{fig:hires_point}.}
  \begin{center}
  \begin{tabular}{c | c c c c c c c c c c}
    \hline
    \hline
    Bin & 1 & 2 & 3 & 4 & 5 & 6 & 7 & 8 & 9 & 10 \\
    \hline
    A & 67.9 & 55.3 & 45.5 & 36.9 & 28.8 &  20.7 &  12.3 &   3.3 & -12.1 & -32.0 \\
    B &  2.0 &  3.0 &  3.8 &  4.7 &  5.5 &   6.6 &   7.9 &   9.4 &  17.6 &   0.0 \\
    C & -3.1 & -4.4 & -5.6 & -7.0 & -8.8 & -11.5 & -15.7 & -26.2 & -19.1 &   0.0 \\
    \hline
    \hline
  \end{tabular}
  \end{center}
  \label{tab:exposure}
\end{table}

\begin{figure}
  \begin{center}
    \includegraphics[width=4in]{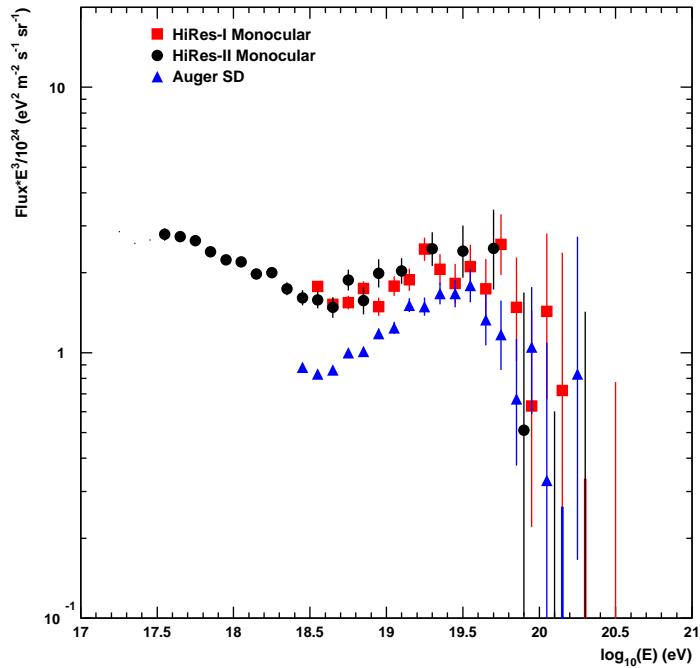}
  \end{center}
  \caption{\label{fig:energy_scale}
Energy spectrum [$E^3J$] for HiRes-1 and HiRes-2 monocular data 
\cite{Abbasi_PRL_2008} and for the surface detector data from the Pierre 
Auger Observatory \cite{Perrone_ICRC2007}.}
\end{figure}

Figure \ref{fig:energy_scale} shows the monocular spectra for the two HiRes
sites \cite{Abbasi_PRL_2008} and that of the Pierre Auger Observatory 
\cite{Perrone_ICRC2007}.
At the highest energies where Auger observes an anisotropy signal, the energy 
scales of HiRes and Auger differ by about 10\%.
To account for this difference, the energy scale of the HiRes stereo data set
used in this analysis has been decreased by 10\% to agree with the
Auger energy scale.
All energies quoted for the HiRes data from this point on will include 
this 10\% shift.
There are 13 events with energies greater than 56~EeV in the full HiRes stereo 
data set, the same number as in the Auger test data set.

\section{The V{\'e}ron-Cetty and V{\'e}ron catalog}

In this paper, we report on searches for correlations between the pointing
directions of ultrahigh energy cosmic rays observed stereoscopically by the
HiRes experiment and AGN from the V{\'e}ron-Cetty and V{\'e}ron (VCV) catalog, 
12$^{th}$ edition \cite{VeronCetty.Veron_AA_2006}.
The VCV catalog includes 
$\sim 22000$ AGN, $\sim 550$ BL-Lacs and $\sim 85000$ quasars 
compiled from observations made by other scientists, and does not evenly 
cover the sky.
Not only does the Galaxy and its associated dust cover large parts of the sky, 
particularly in the southern hemisphere, making the identification of AGN 
extremely difficult in those areas, but some of the sky surveys included in 
the catalog have covered only small bands of the sky.
This makes the total density of AGN in the VCV catalog very uneven across 
the sky in a way that is neither totally random nor systematic.
The locations of a closer subset of sources, with redshift $z < 0.1$,
are more evenly distributed.

\begin{figure}
  \begin{center}
    \includegraphics[width=4in]{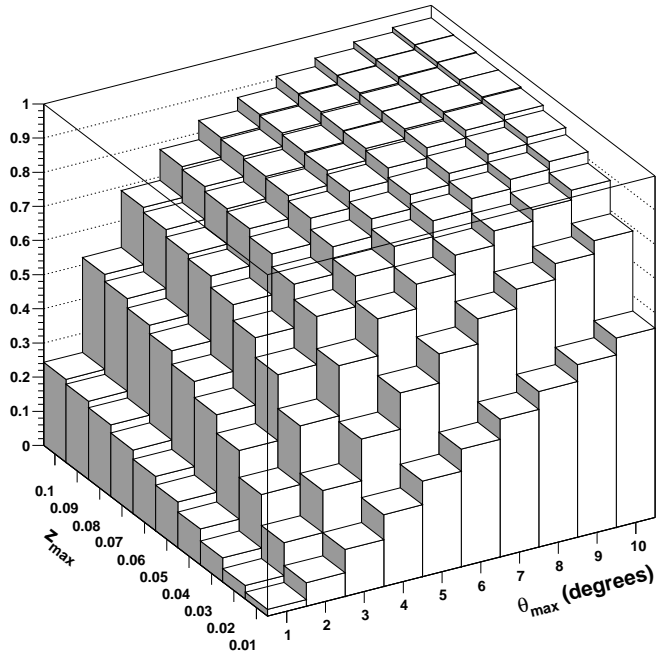}
  \end{center}
  \caption{\label{fig:mcave_1975}
The average fraction of correlated events found in 5000 simulated 
sets of isotropic events with identical statistics to the HiRes data for 
$E > 56$~EeV as a function of $\theta_{max}$ and $z_{max}$.
The fraction of correlated pairs of simulated events with AGN is 0.02 at 
($1.0^\circ$, 0.010); 96\% of events are correlated at ($10.0^\circ$, 0.100).
}
\end{figure}

One property of the search method in ($\theta_{max}$, $E_{min}$, $z_{max}$) 
is that the large size of the catalog and the size of the correlation angle 
circles determine that one can scan over only a narrow range of 
$\theta_{max}$ and $z_{max}$.
To illustrate this using simulated events with isotropically distributed 
pointing directions, Figure \ref{fig:mcave_1975} shows 
that the number of random pairings with AGN is determined by the choice of 
$\theta_{max}$ and $z_{max}$.
As $\theta_{max}$ and $z_{max}$ are increased, the number of random 
pairings increases, rapidly overcoming any real correlations between cosmic 
rays and AGN.

\section{Method}

We perform three searches for correlations between cosmic rays and AGN. 
In the first search we look for correlations in the HiRes stereo data
using the ($\theta _{max}$, $E_{min}$, $z_{max}$) parameters prescribed by 
the Auger collaboration \cite{Auger_AGN_2007}.
In the second, we divide our stereo data into two equal parts in a random 
manner, determine the optimum search parameters in the first half of the 
data by scanning in a three-dimensional grid in 
($\theta _{max}$, $E_{min}$, $z_{max}$),
and then examine the second half of the data using these ``optimum'' 
parameters.
By choosing the best parameters from the first half of the data and using
them to form a hypothesis to be tested using a statistically independent
sample, no statistical penalties are incurred in the application to the 
second half of the data.
In the third and last search, we analyzed the complete data set using 
the statistical prescription described by Finley and Westerhoff
\cite{Finley.Westerhoff_AstropartPhys_2004}
(see also Tinyakov and Tkachev \cite{Tinyakov.Tkachev_PhysRevD_2004})
to arrive at a chance probability that includes the statistical penalty 
from scanning over the entire data set.
Finally, in addition to searching for correlations with AGN, we analyzed 
the degree of auto-correlation in the stereo data over all possible angles 
and values of $E_{min}$.

To arrive at the appropriate chance probabilities for the numbers of 
correlations seen in each method, we generated 5001 random samples of events 
using the hour angle - declination method 
\cite{Atkins_ApJ_2003,Ivanov_ICRC2007}.
In this method the hour angle and declination of one event and 
the sidereal time of another are randomly paired to generate a sky plot 
with the same number of events as the data.
Such a sample reproduces the overall observed distribution of events very well.

\subsection{Search for Correlations using the Auger criteria}

The Auger collaboration has reported the results of searches in
($\theta _{max}$, $E_{min}$, $z_{max}$) over two independent data sets.
In a scan over the first data set, 12 of the 15 events with 
$E_{min}~=~56.0$~EeV were found to lie within $\theta _{max}~=~3.1^\circ$ of 
AGN with $z_{max}~=~0.018$ with 3.2 chance pairings expected.
Using the parameters ($3.1^\circ$, 56.0 EeV, 0.018), 8 of 13 events in an 
independent test data set were found to be paired with AGN with 2.7
chance pairings expected.
The chance probability for this occurrence was found to be 0.0017
\cite{Auger_AGN_2007,Auger_AGN_2007_full}.

\begin{figure}
  \begin{center}
    \includegraphics[width=6in]{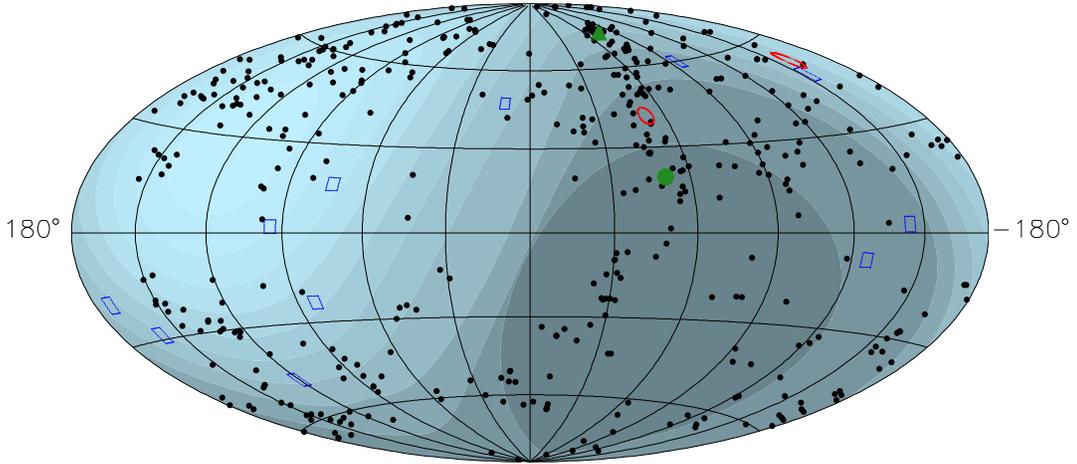}
  \end{center}
  \caption{\label{fig:auger_point}
Sky map in Galactic coordinates.
The black dots are the locations of the 457 AGN and 14 QSOs with
redshift $z<0.018$.
The green circle and triangle mark the locations of Centaurus A and M87, 
respectively.
The red circles (with radii of $3.1^\circ$) mark the 2 correlated events.
The blue squares mark the locations of the 11 uncorrelated events.
Of the eleven blue shaded regions, the 10 lightest shades delineate 
regions of constant exposure in HiRes as given in Table \ref{tab:exposure}.
The darkest shade indicates the region with no exposure.}
\end{figure}

A scan of the entire HiRes data set at ($3.1^\circ$, 56.0 EeV, 0.018) found
2 AGN pairings for a total of 13 events.
Figure \ref{fig:auger_point} shows the locations of the 2 correlated events
and the 11 uncorrelated events.
We looked for correlations in the 5000 simulated data sets at  
($3.1^\circ$, 56.0 EeV, 0.018)
and found the average number of correlated pairs to be 3.2.
In addition, 4121 sets had 2 or more correlated events for a chance 
probability of 82\%.
We thus find no evidence for correlations of cosmic-ray events with AGN 
in our field of view at ($3.1^\circ$, 56.0 EeV, 0.018).
The HiRes data are therefore consistent with random correlations.

\subsection{Search in two independent data sets}

Next, we randomly divide the HiRes stereo data into two equal sets, first 
examining only one half and setting the other aside.
We scan the first half simultaneously 
in $\theta_{max}$ from 0.1 to $4.0^\circ$ in bins of $0.1^\circ$, 
in $E_{min}$ from $10^{19.05}$ to $10^{19.80}$~eV in bins of 0.05 decade, 
and with an AGN $z_{max}$ from 0.010 to 0.030 in bins of 0.001.
For each grid point in the scan, the total number of cosmic rays correlated 
with at least one AGN is accumulated.
We then conduct the same scan in each of 5000 simulated sets with identical 
statistics to the first half, adding up the total number of correlations in
each set for each grid point.
At each point, the number of correlated events in each of the 5000 simulated
sets is compared with the result in the first half of the data.
The criteria for the most significant correlation were found to be 
($1.7^\circ$,~15.8~EeV,~0.020) with 20 correlated events from a total of 97.
Only 25 of 5000 simulated sets had 20 or more correlations.

Using these criteria as our hypothesis, we then examine the second half of the 
data at ($1.7^\circ$, 15.8 EeV, 0.020) and find 14 correlated pairs from 
101 events.
In a set of 5000 simulated events with identical statistics to the second
half, 741 sets contained 14 or more correlated events for a chance 
probability of 15\%.
For comparison, the point with the most significant correlation in the 
second half occurs at ($2.0^\circ$, 20.0 EeV, 0.016) with 14 correlated 
events of a total 69 and a chance probability of 1.5\%.
These results are again consistent with random correlations.

\subsection{\label{sec:fw_method} Scanning the entire data set}

We follow the prescription of Finley and Westerhoff 
\cite{Finley.Westerhoff_AstropartPhys_2004}
for determining the most significant correlation in the entire data set 
while also
calculating an appropriate statistical penalty for scanning over the entire
data set.
We scan the data simultaneously in $\theta _{max}$, $E_{min}$ and $z_{max}$
counting the number of correlated events, $n_{corr}$ at each point.
This process is repeated for each of the 5001 simulated sets with $P_{data}$, 
the probability for observing $n_{corr}$ or more correlations at 
($\theta _{max}$,~$E_{min}$,~$z_{max}$) calculated from

\begin{equation}
  \label{eq:pdata}
  P_{data}(\theta_{max}, z_{max}, E_{min}) =
  \sum _{n=n_{corr}} ^{\infty} P_{mc}(\theta_{max}, z_{max}, E_{min}, n)
\end{equation}

\noindent
where $P_{mc}(\theta_{max}, z_{max}, E_{min}, n)$ is the fraction of the
first 5000 simulated sets with exactly $n$ events at 
($\theta _{max}$,~$E_{min}$,~$z_{max}$).
The value of $P_{min}$ is then taken to be the values of
($\theta _c$, $E_c$, $z_c$) which minimize $P_{data}$.
This is found to occur at the critical values 
($2.0^\circ$,~15.8~EeV,~0.016) where there are 36 correlated events out 
of 198
in the data and 9 of 5000 simulated sets with 36 or more correlated 
events, for a chance probability of 0.18\%.

To find the true significance of this signal, we apply the same process to
each of the first 5000 simulated sets, finding the value 
$P_{min}^i~=~P^i(\theta _c ^i,~E_c^i,z_c^i)$ by comparing $n_{corr}^i$ with
$n_{corr}$ for the other 5000 sets.
We then count the number of simulated sets $n_{mc}^*$ for which
$P_{min}^i \leq P_{min}$.
The chance probability is then found as

\begin{equation}
  \label{eq:pchance}
  P_{chance} = \frac{n_{mc}^*}{5000}.
\end{equation}

\begin{figure}
  \begin{center}
    \includegraphics[width=6in]{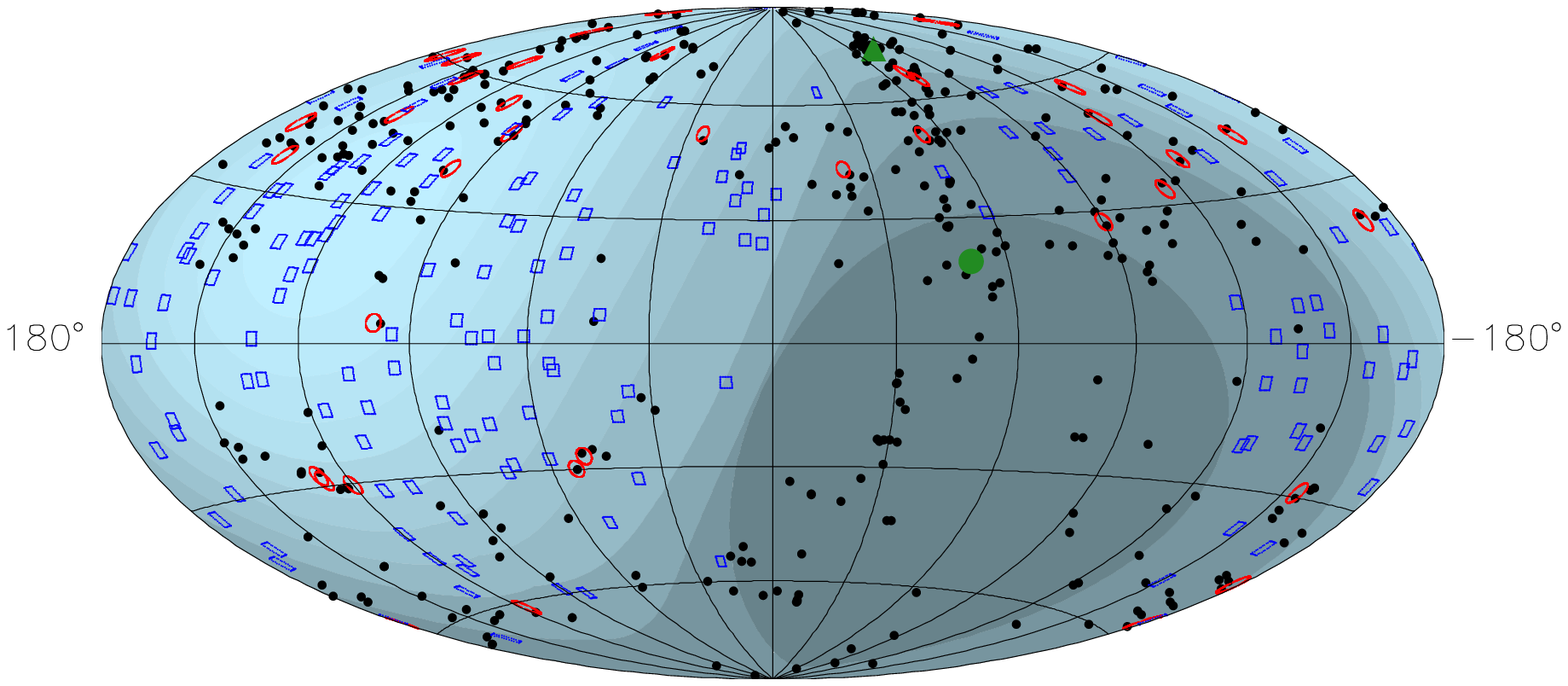}
  \end{center}
  \caption{\label{fig:hires_point}
Sky map in Galactic coordinates.
The black dots are the locations of the 389 AGN and 14 QSOs with
redshift $z<0.016$.
The green circle and triangle mark the locations of Centaurus A and M87, 
respectively.
The red circles (with radii of $2.0^\circ$) mark the 36 correlated 
events at ($2.0^\circ$, 15.8 EeV, 0.016).
The blue squares mark the locations of the 162 uncorrelated events.
Of the eleven blue shaded regions, the 10 lightest shades delineate 
regions of constant exposure in HiRes as given in Table \ref{tab:exposure}.
The darkest shade indicates the region with no exposure.}
\end{figure}

\noindent
In this, our most robust method, there were 1210 simulated sets with 
$P_{min}^i$ values of 0.0018 or less for a chance probability, 
$P_{chance} = 24$\%.
Figure \ref{fig:hires_point} shows a sky map of the most significant 
correlation in the HiRes data.
From this final analysis, we draw the same conclusion: HiRes data are 
consistent with random correlations with AGN.

\section{Auto-correlation analysis}

\begin{figure}
  \begin{center}
    \includegraphics[width=4in]{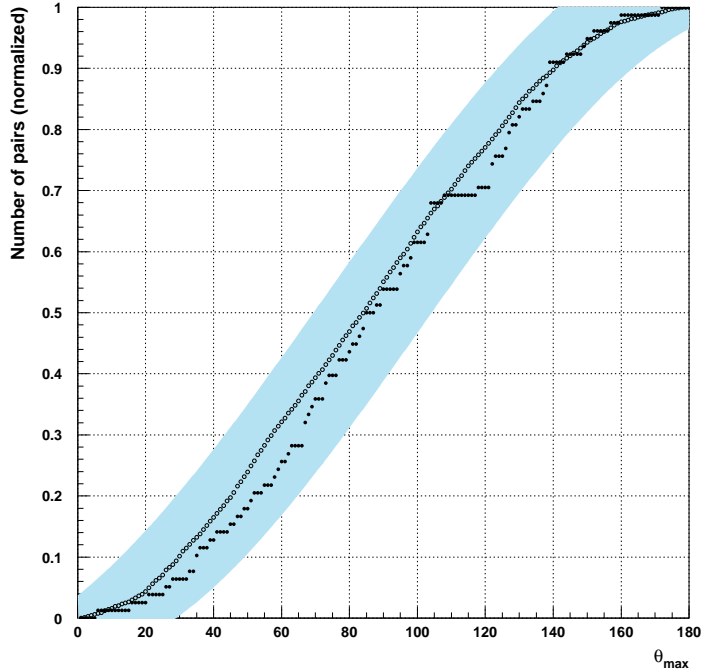}
  \end{center}
  \caption{\label{fig:autocorrelation}
Normalized number of pairs as a function of $\theta_{max}$.
The 13 events above 56 EeV in the HiRes data are shown in closed circles.
The open circles are the average of 2000 simulated sets.
The gray shaded region represents the 1$\sigma$ uncertainty in the
distribution of simulated sets.}
\end{figure}

In addition to searching for correlations with AGN, studies of
auto-correlation can be useful for searching for anisotropy in the data.
We have analyzed the degree of auto-correlation in the data over all possible
angles and made comparisons with the average number of pairs of events 
for 2000 isotropic simulated data sets.
We find no evidence of auto-correlation for any values of $E_{min}$.
Figure \ref{fig:autocorrelation} shows a comparison of the normalized number 
of pairs of events with energies above 56~EeV in the stereo data to the 
average normalized number of pairs for 2000 isotropic simulated data sets.
The 1$\sigma$ uncertainty is found by ordering the simulated sets by their
maximum deviation from the average and plotting only the first 68\% of 
those simulated sets.

As a further check, we scan the data in $\theta_{max}$ and $E_{min}$ and 
determine a statistical penalty using the same method presented in
Section 3.3.
We scan the data in $\theta_{max}$ from $0.5^\circ$ to $30.0^\circ$ in bins of
$0.5^\circ$ and in $E_{min}$ from $10^{19.05}$ to $10^{19.80}$ eV in
bins of 0.05 decade.
The critical values which minimize $P_{data}$ are found to occur at
($2.0^\circ$, 44.7 EeV) where there is one pair of events out of a 
possible 406 in the data and 227 of 1000 simulated sets with one or more 
pairs for a chance probability of 23\%.
Applying the same process to the 1000 simulated sets, we find 971 sets 
for which the critical point occurs with a chance probability less than 
23\%.
The probability of measuring the observed degree of correlation in an isotropic
data set is 97\%.

\section{Conclusions}

We have searched for correlations between the pointing directions of HiRes
stereo events with AGN from the the V{\'e}ron-Cetty V{\'e}ron catalog using
three different methods.
As search parameters for our analysis, we used the maximum difference in
angle between the cosmic-ray pointing direction and an AGN $\theta _{max}$,
the minimum cosmic-ray energy $E_{min}$, and the maximum AGN redshift 
$z_{max}$.

Our first analysis, using the criteria prescribed by the Pierre Auger 
Observatory for their most significant correlation, ($3.1^\circ$, 56.0 
EeV, 0.018), finds 2 correlated of 13 total events with an expectation 
of 3.2 chance correlations.
The corresponding chance probability was found to be 82\%.

In our second search the total HiRes stereo data were then divided into two 
equal but random parts and we performed a scan in $\theta _{max}$, $E_{min}$ 
and $z_{max}$ over one half of the data to determine which parameters 
optimized the correlation signal.
We then examined the other half of the data using these search parameters and 
found a smaller signal with a chance probability of 15\%.

Finally, we examined the entire HiRes stereo data using a more robust method 
to calculate the chance probability with appropriate statistical penalties.
The most significant correlation was found to occur at 
($2.0^\circ$, 15.8 EeV, 0.016) 
with 36 correlated of 198 total events.
This corresponds to a chance probability of 24\%.

We conclude that there are no significant correlations between the HiRes stereo
data and the AGN in the V{\'e}ron-Cetty V{\'e}ron catalog.
We also examined the degree of auto-correlation at all angles and energies.
The probability that the data are consistent with isotropy is 97\%.

\section*{Acknowledgments}
This work was supported by US NSF grants PHY-9100221, PHY-9321949,
PHY-9322298, PHY-9904048, PHY-9974537, PHY-0073057, PHY-0098826,
PHY-0140688, PHY-0245428, PHY-0305516, PHY-0307098, PHY-0649681, and
PHY-0703893, and by the DOE grant FG03-92ER40732.  We gratefully
acknowledge the contributions from the technical staffs of our home
institutions. The cooperation of Colonels E.~Fischer, G.~Harter and
G.~Olsen, the US Army, and the Dugway Proving Ground staff is greatly
appreciated.

\bibliography{references}
\bibliographystyle{elsart-num}

\end{document}